\documentclass{ws-procs9x6}
\def\itmb{\begin{itemize}}
\def\itme{\end{itemize}}
\def\enmb{\begin{enumerate}}
\def\enme{\end{enumerate}}
\def\eqnb{\begin{equation}}
\def\eqne{\end{equation}}
\def\eqab{\begin{eqnarray}}
\def\eqae{\end{eqnarray}}
\def\dis{\displaystyle}

\usepackage[dvips]{color}
\usepackage{epsfig}

\begin{document}
\title {Color Confinement in Lattice Landau Gauge } 
\author{
Hideo Nakajima}
\address{
Department of Information Science, Utsunomiya University \\
E-mail: nakajima@is.utsunomiya-u.ac.jp}
\author{
Sadataka Furui }
\address{
School of Science and Engineering, Teikyo University\\
E-mail: furui@umb.teikyo-u.ac.jp}

\maketitle

\abstracts{The problem of color confinement in lattice Landau gauge QCD is 
briefly reviewed in the light of Kugo-Ojima criterion and 
Gribov-Zwanziger theory. 
The infrared properties of the Landau gauge QCD is studied by the
measurement of the gluon propagator, ghost propagator, Kugo-Ojima parameter 
and the running coupling via $\beta=6.0, 16^4, 24^4, 32^4$ 
and $\beta=6.4, 32^4, 48^4$ lattice simulation. 
The data are analyzed in the principle of minimum sensitivity(PMS) 
in the $\widetilde{MOM}$ scheme. We observe the running 
coupling $\alpha_s(q)$ has a maximum about 1 and decreases 
near $q=0$. The Kugo-Ojima parameter, which is expected to 
be 1 was about 0.8. }

\section{Introduction}

In 1978 Gribov proposed that the integration range of the gauge field should be
restricted to the region where its Faddeev-Popov determinants are positive and
that the restriction will cancel the infrared singularity of the gluon propagator and results in the singularity of the ghost propagator, which makes the linear potential between colored sources, and the color confinement will be realized\cite{Gv}.
Almost at the same time, Kugo and Ojima proposed color confinement criterion based on the 
BRST(Becchi-Rouet-Stora-Tyutin) symmetry without Gribov's problem taken into
account\cite{KO}.

Kugo-Ojima two-point function is defined in the lattice simulation as
\eqnb
(\delta_{\mu\nu}-{p_\mu p_\nu\over p^2})u^{ab}(p^2)=
\dis{1\over V}
\sum_{x,y} e^{-ip(x-y)} \left \langle  {\rm tr}({\lambda^a}^{\dag}
D_\mu \dis{1\over -\partial D}[A_\nu,\lambda^b] )_{xy}
\right \rangle
\eqne
 where $u^{ab}(0)=\delta^{ab}u(0)$ and 
$ 1+u(0)=Z_1/Z_3=1/\tilde Z_3$.  $Z_3$ and $\tilde Z_3$ are gluon- and ghost- wave function renormalization factor, respectively and $Z_1$ is the gluon vertex renormalization factor. 
Kugo expressed the ghost propagator $G(p^2)$ in the infrared asymptotic 
region as\cite{Ku} $G(q^2)\sim \dis{1\over q^2(1+u(0)+O(q^2))}$.

In 1994, Zwanziger analyzed various regions of the Landau gauge in connection
with Gribov's problem, and developed the formulation of lattice Landau 
gauge\cite{Zw}. 
From $L$-periodic link variables, $U_{x,\mu}\in SU(n)\ (n=3)$, 
we define gauge field $A_{\mu}(U)$, and then there are 
two possible options of $A_{\mu}(U)$, $U-$linear (Zwanziger's) type 
$A_{x,\mu}=(U_{x,\mu}-U_{x,\mu}^{\dag})/2|_{traceless\ part}$,
and $\log\ U$ (ours)\cite{NF} type $U_{x,\mu}=e^{A_{x,\mu}}$. Here brief 
review is given below in terms of common language to both options of the 
definition.
In each definition, $\delta A_\mu$ under infinitesimal gauge 
transformation $g=e^{\epsilon}$ is given as 
$\delta A_\mu^g=D_\mu (U)\epsilon$ with the covariant derivative defined as
$
D_{\mu}(U)\phi=S(U_\mu)\partial_{\mu}\phi+[A_\mu,{\overline {\phi}}^\mu],
$ 
where operations $\partial_\mu$ and ${\bar {\ }}^\mu$ on the scalar 
$\phi$ to give vectors are defined as 
$(\partial_\mu \phi)_{x,\mu}=\phi(x+\mu)-\phi(x)$ and $
({{\overline \phi}^\mu})_{x,\mu}=\left(\phi(x+\mu)+\phi(x)\right)/2$, 
respectively, 
and the operation $S(U_\mu)$ on a vector $B_\mu$ is defined 
for each option as 
$
S(U_\mu)B_\mu=(1/ 2)\left.\left\{ (U_{x,\mu}+U_{x,\mu}^\dag)/2
,B_{x,\mu}\right\}\right |_{traceless\ part}
$ 
in $U$-linear definition, 
$
S(U_\mu)B_\mu=S({\mathcal A}_{x,\mu})B_{x,\mu}
=\{({\mathcal A}_{x,\mu}2)/{\rm th}({\mathcal A}_{x,\mu}/2)\}B_{x,\mu}
$ 
in $\log U$ definition with 
${\mathcal  A}_{x,\mu}=adj_{A_{x,\mu}}=[A_{x,\mu},\ \cdot\ ]$.

The Landau gauge $\partial A=0$ can be characterized\cite{MN} 
such that $\delta F_U(g)=0$ for any $\delta g$, in use of the optimizing 
functions $F_U(g)$; $
F_U(g)=\sum_{x,\mu}{\rm tr} \left\{2- (U^g_{x,\mu}+{U^g_{x,\mu}}^\dag)\right\}
$ in $U$-linear definition, and $
F_U(g)=\sum_{x,\mu}{\rm tr} \left({{A^g}_{x,\mu}}^{\dag}\ A^g_{x,\mu}\right)
\equiv \langle A^g_\mu|A^g_\mu \rangle$ in $\log U$ definition.
The variation
$\Delta F_U(g)$, under infinitesimal gauge 
transformation $g^{-1}\delta g=\epsilon$, reads as
$\Delta F_U(g)=-2\langle \partial A^g|\epsilon\rangle+
\langle \epsilon|-\partial { D(A^g)}|\epsilon\rangle+\cdots$.
The {\bf fundamental modular region} is specified by the
{\bf global minimum} along the gauge orbits, i.e., 
$
\Lambda_L=\{U|\ A=A(U),\ F_{U}(1)={\rm Min}_gF_{U}(g)\}
$, 
$
\Lambda_L\subset \Omega_L
$, 
where $\Omega_L$ is called as {\bf Gribov region} (local minima), and 
$
\Omega_L=\{U|-\partial { D(U)}\ge 0\ ,\ \partial A=0\}.
$
So far, all field variables are supposed to be $L$-periodic.
Zwanziger further defined the {\bf core region} $\Xi_L$ as a set of 
the global minimum points
of $F_U(g)$ in the extended gauge transformation $g=e^\omega e^{\theta x}$
where $U$ and $\omega$ are $L$-periodic, and constant $\theta_\mu$'s belong to 
an arbitrary Cartan subalgebra, i.e., $[\theta_\mu,\theta_\nu]=0$: 
$
\Xi_L=\{U|\ F_U(1)={\rm Min}_{g=e^\omega e^{\theta x}} F_U(g) \}\subset 
\Lambda_L.
$
Putting $g=e^\omega e^{\theta x}$ with arbitrary $\omega$ and $\theta_\mu$ of order $\epsilon$, 
one has the variation of $F_U(g)$ up to $O(\epsilon^2)$ for $U\in \Xi_L$ as 
$
\Delta F_U(g)=2\langle A_\mu|\theta_\mu \rangle
+\langle  \omega' | -\partial D|\omega'\rangle
-\langle  D\theta|(-\partial D)^{-1}|D\theta\rangle
+\langle \theta_\mu|S(U_\mu)| \theta_\mu \rangle\ge 0,
$
 where $\omega'=\omega-(-\partial D)^{-1}D\theta$.
The non-negative sum of the third and fourth terms is expressed by 
$
-\langle \theta_\mu|H_{\mu\nu}| \theta_\nu \rangle\ 
$
with the {\bf horizon tensor} 
$
H_{\mu\nu}=-D_\mu(-\partial D)^{-1}D_\nu
-\delta_{\mu\nu}S(U_\mu).
$
Taking the trace of the operator $H_{\mu\nu}$ with respect to the normalized
constant colored vectors $\eta_\mu^{\nu,a}=\delta_{\mu\nu}\lambda^a$ with 
${\rm tr}\lambda^{a\dag}\lambda^b=\delta_{ab}$, 
one defines the {\bf horizon function} $H(U)$ as 
$
H(U)=\sum_{\nu,a}\langle \eta_\mu^{\nu,a}|H_{\mu\rho}|
\eta_\rho^{\nu,a}\rangle
=
\sum_{\mu,a}\langle \eta_\mu^{\mu,a}|
-D_\mu(-\partial D)^{-1}D_\mu
|\eta_\mu^{\mu,a}\rangle
-(n^2-1)E(U)
\equiv h(U)V\ 
$
where $(n^2-1)E(U)=\sum_{x,\mu}{\rm tr}(\lambda^{a\dag} S(U_{x,\mu})\lambda^a)$. 
Thus one has for $U\in \Xi_L$ that $\overline {A_\mu}=V^{-1}
\sum_x A_{x,\mu}=0$ and $H(U)\le 0$, where $V=L^4$. 
Zwanziger hypothesized that {\em the dynamics on $\Xi_L$ tends to
that on $\Lambda_L$ in the infinite volume limit}, and derived 
the {\bf horizon condition}, statistical 
average $\bigl \langle h(U) \bigr \rangle=0$, in the infinite volume limit.
Taking the Fourier transform of the tensor propagator of the color point source, 
$\bigl \langle -D_\mu(-\partial D)^{-1}D_\nu \bigr \rangle_{xa,yb}$, one has 
$
G_{\mu\nu}(p)\delta^{ab}=
(e/d)(p_\mu p_\nu/ p^2)\delta^{ab}
-\{\delta_{\mu\nu}-(p_\mu p_\nu/ p^2)\}u^{ab}(p^2)
$
, where $e=\langle E(U)\rangle/V$ and dimension $d=4$.
Putting Kugo-Ojima parameter as $u(0)=-c$ and comparing $\lim_{p_\mu\to +0}
G_{\mu\mu}$ with $\bigl \langle h(U) \bigr \rangle=0$ one finds that the horizon condition 
reduces to
\eqnb
\left \langle {h(U)\over n^2-1}\right \rangle
=\left(\dis{e\over d}\right)+(d-1)c-e=(d-1)\left(c-\dis{e\over d}\right)
\equiv (d-1)h=0.
\eqne
Kugo-Ojima's and Zwanziger's arguments emerge to be consistent
with each other provided the lattice covariant derivative meets with 
the continuum one $e/d=1$. Accordingly, Zwanziger derived independently 
the same characteristic singular behavior of the ghost propagator\cite{Zw} 
as Kugo's\cite{Ku}, both perturbatively in the sense that the 
diagrammatic expansion was used. 
Recent Dyson-Schwinger analyses\cite{FAR} 
and numerical simulations are to be noticed in this respect as well.
It should be pointed out here that our numerical simulation does not pursue 
the selective core region dynamics, 
but rather checks the Zwanziger hypothesis in our standard method\cite{NFY1}.  
\section{The gluon propagator}
The gluon propagator and its dressing function are defined numerically as
\eqnb
D_{\mu\nu}(q)={V^{-1}\over n^2-1}\sum_{x,y}e^{-iq(x-y)}
{\rm tr}\bigl\langle A_\mu(x)^\dagger A_\nu(y)\bigr\rangle
=(\delta_{\mu\nu}-{q_\mu q_\nu\over q^2})D_A(q^2)
\eqne
and $Z(q^2)=q^2 D_A(q^2)$, respectively.
The infrared behavior of $Z(q^2)$ is
parametrized as $(q^{-2})^{\alpha_D}$, 
and we observed that $\alpha_D<0$ for $L\geq 24$.

As a kinematical structure of the Gribov region, 
Zwanziger gave a bound\cite{Zw1}  
on the constant mode 
$\overline {A_\mu}=V^{-1}\sum_{x}A_{x,\mu}$ for
$A\in \Omega_L$, as
$|\overline {A_\mu}|\le \dis{2\over \alpha}\{\tan (\dis{\pi\over L})\}
\equiv \sigma(L)$ 
with a lattice size $L$, $V=L^4$, and 
$\alpha=\sqrt{3/2}$ in $SU(3)$ and 
$\alpha=\sqrt{2}$ in $SU(2)$, respectively. 
This fact can be derived irrespective of the
gauge field options. 
The Zwanziger's bound was checked to hold 
in case $\beta=6.4$, $L=48$, as 
$|\overline{A_\mu}|\leq 0.005$ vs. $\sigma(L)= 0.10$ for each sample.
The propagator at zero momentum 
has an extra factor $V$, as  $D_A(0)\sim V{\overline A}^2$. 
Although $\displaystyle {\sqrt V} 
{\bigl \langle} \overline {A_\mu} {\bigr\rangle} =0.23\pm 4.2$, 
the propagator has the value 
$3D_A(0)=\frac{3}{4\cdot 8} V {\bigl \langle} \sum_{\mu,a} 
{{\overline{A_\mu}}^a}^2\bigr \rangle \simeq 150\pm 30$ 
(averaged over samples). 
It implies that the gluon propagator is infrared finite. 
Since there is a strong cancellation between samples, the sample average is
$\frac{3}{4} V \sum_\mu \langle {\overline {A_\mu}} \rangle^2=0.16\pm 31$.

Here we review Zwanziger's strong argument for the vanishing of zero 
momentum connected Green functions in the infinite volume limit\cite{Zw1}. 
Let a partition function be 
$
Z(J)=\int dA \rho (A) e^{J\cdot A},
$
and for constant $J_x=h$, it reads that 
$
Z(h)=\int dA \rho (A) e^{h\cdot {\widetilde A}(0)}=
\int dA \rho (A) e^{Vh\cdot \overline A}
$
, where ${\widetilde {A_\mu}}(0)= a^d \sum_x A_\mu(x)$. 
Let $W(h)=\log Z(h)$, and then the connected Green function 
$V G_c^{(n)}(0,...,0)={\bigl \langle} ({\widetilde A}(0))^n {\bigr \rangle}|_c$ 
is given as 
$
G_c^{(n)}(0,...,0)=\dis{1\over V}\dis{\partial ^n\over \partial h^n}W(h)
$
. 
Note from positivity of $\rho(A)$ 
that $e^{-Vh\sigma(L)}\le Z(h)\le e^{Vh \sigma(L)}$.
Putting $w(h)\equiv W(h)/V$, we obtain
$|w(h)|\le h\sigma(L) \to 0$ 
as $L\to \infty$. Hence it proves that $G_c^{(n)}(0,...,0) \to 0$ in the 
infinite volume limit. However, it should be noted that this infinite volume
limit is only proved directly for connected Green functions. 

\begin{figure}[htb]
\begin{minipage}[b]{0.47\linewidth}
\begin{center}
\epsfysize=100pt\epsfbox{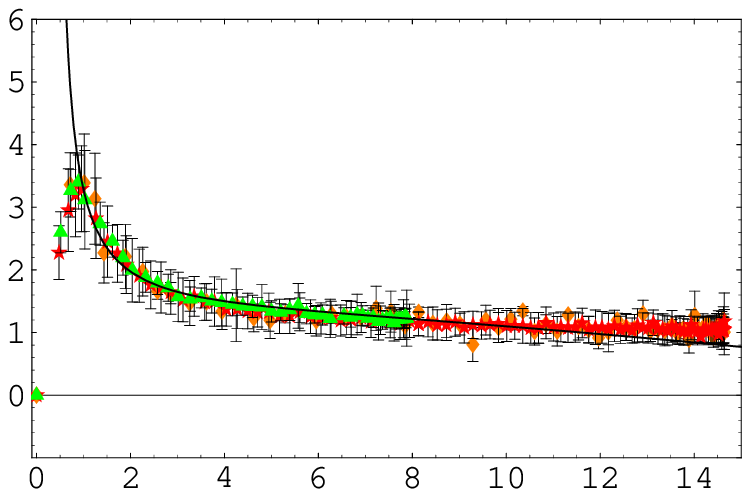}
\caption{ The gluon dressing function as the function of the momentum $q(GeV)$. $\beta=6.0$, $24^4, 32^4$ 
and $\beta=6.4$, $48^4$ in $\log U$ version. The fitted line is that of PMS in $\widetilde{MOM}$ method. \label{gl243248}}
\end{center}
\end{minipage}
\hfil
\begin{minipage}[b]{0.47\linewidth}
\begin{center}
\epsfysize=100pt\epsfbox{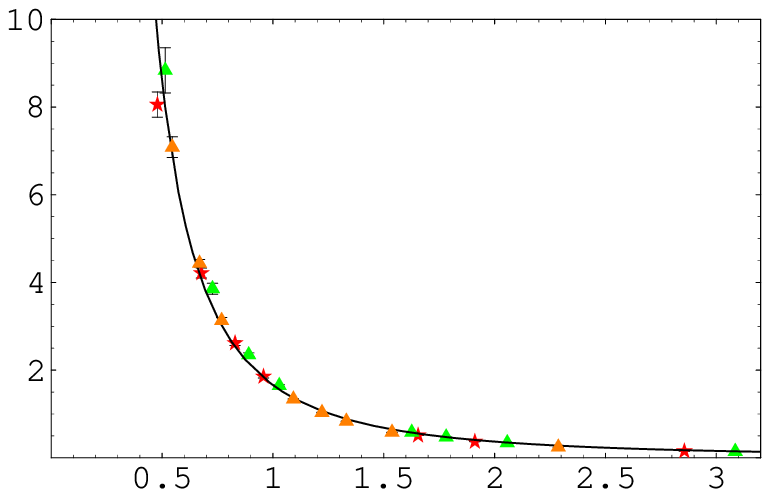}
\caption{ The ghost propagator as the function of the momentum $q(GeV)$. $\beta=6.0$, $24^4, 32^4$ and $\beta=6.4$, $48^4$ in $\log U$ version.
 The fitted line is that of PMS in $\widetilde{MOM}$ method.\label{ghost}}
\end{center}
\end{minipage}
\end{figure}

In Fig\ref{gl243248}, we plot the gluon dressing function of
$24^4, 32^4$ and $48^4$ lattices. The infrared properties of the gluon dressing
function is studied in the principle of minimum sensitivity(PMS)\cite{pms,vanacol} in 
$\widetilde{MOM}$ scheme. The parameter $y=0.02227$ is fixed at $q=1.97GeV$ 
and $q^2D_A(q^2)=Z(q^2,y)|_{y=0.02227}$ is compared with the lattice data.

\section{The ghost propagator}
In use of color
source $|\lambda^a x\rangle$ normalized as
${\rm tr} \langle \lambda^a x|\lambda^b x_0\rangle=\delta^{ab}\delta_{x,x_0}$,
the ghost propagator is given 
by the Fourier transform of 
\eqnb
D_G^{ab}(x,y)=\left \langle {\rm tr} \langle \lambda^a x|\{-\partial D(U)\}^{-1}
|\lambda^b y\rangle \right \rangle
\eqne
where the outmost $\bigl \langle \bigr \rangle$ denotes an average over 
samples $U$. 
The ghost dressing function is defined as $G^{ab}(q^2)=q^2 {D_G}^{ab}(q^2)$,
and in the infrared region, it is parametrized as $(p^{-2})^{\alpha_G}$.
Although the ghost dressing function is monotonic in the case of $L\leq 32$, the lowest momentum point of  $L=48$ is suppressed.  

We compare in Fig.\ref{ghost} the result of the PMS method in $\widetilde{MOM}$ scheme\cite{vanacol}
$D_G(q^2)=-Z_g(q^2,y)/{q^2}|_{y=0.02142}$ and the lattice data
of $\beta=6.0, L=24,32$ and $\beta=6.4, L=48$. 

\section{The Kugo-Ojima parameter}
We directly measured the Kugo-Ojima parameter on $16^4, 24^4, 32^4$ and 
$48^4$ lattices. Results are summarized in Table\ref{tab1}. 
\begin{table}[ph]
\tbl{Lattice size dependence of the Kugo-Ojima parameter $c$, the  trace $e$ 
divided by the dimension $d$ and the deviation parameter from the horizon 
condition $h$. The suffix 1 corresponds to the $U$-linear definition, and 2 
to the $\log U$ definition.\vspace*{1pt}}
{\footnotesize
\begin{tabular}{|c|c|rrr|rrr|}
\hline
{} &{} &{} &{} &{}&{}&{}&{}\\[-1.5ex]
{$\beta$} & {$L$} &{ $c_1$ } & { $e_1/d$ } & { $h_1$ } & { $c_2$ } & { $e_2/d$ } & { $h_2$} \\[1ex]
\hline
{6.0}&{ 16}& { 0.576(79)} &  { 0.860(1)} & { -0.28} &{ 0.628(94)}&{ 0.943(1)} & { -0.32}\\[1ex]
{6.0}&{ 24}& { 0.695(63)}  & { 0.861(1)} & { -0.17} &{ 0.774(76)}&{ 0.944(1)} & { -0.17}\\[1ex]
{6.0}&{ 32}& { 0.706(39)}  & { 0.862(1)} & { -0.15} &{ 0.777(46)}&{ 0.944(1)} & { -0.16}\\[1ex]
\hline
6.4 &32& 0.650(39) & 0.883(1) & -0.23 & 0.700(42)& 0.953(1) & -0.25\\[1ex]
6.4& 48&           &          &       & 0.720(49)& 0.982(1) & -0.26\\[1ex]
\hline
\end{tabular}\label{tab1} }
\vspace*{-13pt}
\end{table}

\begin{figure}[htb]
\begin{minipage}[b]{0.47\linewidth} 
\begin{center}
\epsfysize=100pt\epsfbox{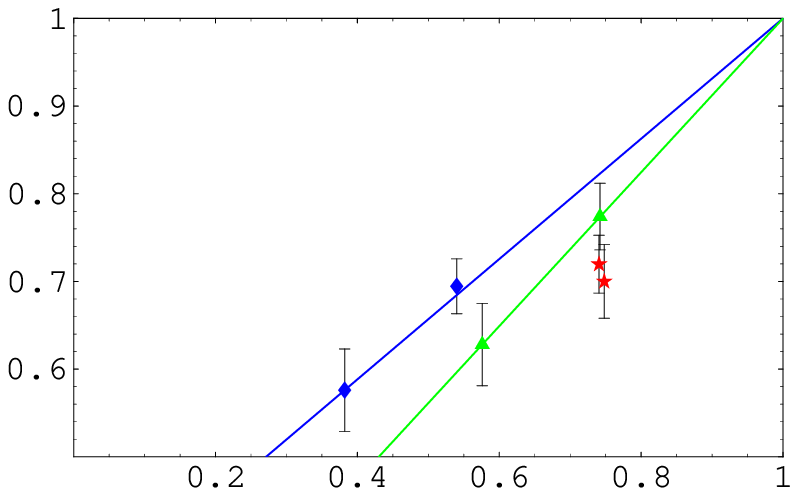}
\end{center}
\caption{ The Kugo-Ojima parameter $c$ as the function of $\log Z_3(1.97GeV)$. 
$\beta=6.0$, $16^4, 24^4$ in $U$-linear(diamond) in $\log U$(triangle) and 
$\beta=6.4$, $32^4, 48^4$ in $\log U$ version (star).\label{kugo}}
\end{minipage}
\hfil
\begin{minipage}[b]{0.47\linewidth} \label{alp3248}
\begin{center}
\epsfysize=100pt\epsfbox{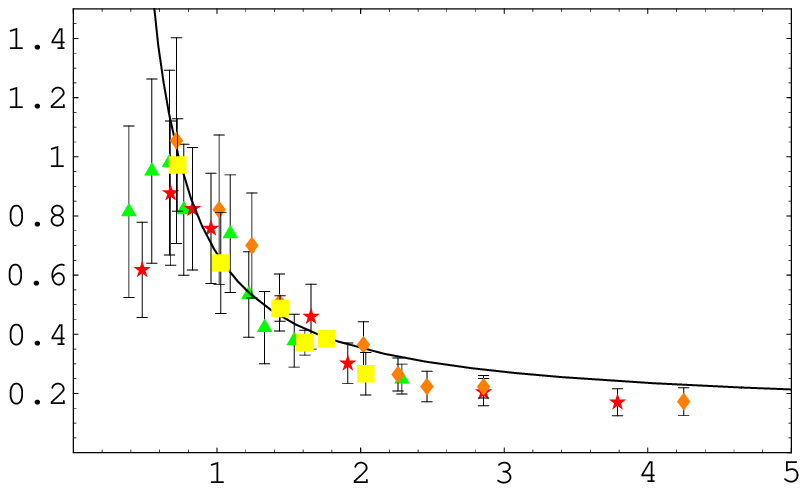}
\end{center}
\caption{The running coupling $\alpha_s(q)$ of $\beta=6.0$, $24^4$(box), 
$32^4$(triangle), $\beta=6.4$, $32^4$(diamond) and $48^4$(star) 
as a function of momentum $q(GeV)$ and the result of the 
PMS method in $\widetilde{MOM}$ scheme.}
\end{minipage}
\end{figure}

In Fig\ref{kugo}, the value $c$ is plotted as a function of $\log Z_3(1.97GeV)$ for  $\beta=6.0$, $L=16,24$ and  $U$-linear and $\log U$ definitions of the gauge field, and for $\beta=6.4$, $L=32,48$ and $\log U$ definition. 
The crossing point of the extrapolation of the points corresponding to 
the $\log U$ and $U-$ linear definitions of the gauge field suggests $c$ in 
the continuum limit $c=1-Z_1/Z_3$ is consistent to 1.
\section{The QCD running coupling}
Using the renormalized gauge field $ A_{r\mu}^a(x)={Z_3}^{-1/2}A_\mu^a(x)$ 
and the ghost field $c_r^b(x)={\tilde Z_3}^{-1/2}c^b(x)$, 
one can show that the vertex renormalization factor of the ghost field 
$\tilde Z_1=Z_gZ_3^{1/2}\tilde Z_3=1$, implies $g^2 Z_3\tilde Z_3^2$ is 
renormalization group invariant.

We estimated the running coupling from the two-point function 
(gluon- and ghost-dressing functions) 
\eqnb
\alpha_s(p^2)=\{g^2/(4\pi)\}Z(p^2){ G(p^2)}^2\simeq 
(pa)^{-2(\alpha_D+2\alpha_G)}.
\eqne

Dyson-Schwinger approach predicts\cite{FAR}
$\alpha_G=0.5953, \alpha_D=-1.1906$ and $\alpha_s(0)=2.972$, while the 
$\widetilde{MOM}$ scheme approach of Orsay group and the PMS 
method\cite{vanacol} predict that $\alpha_s(0)$ is small and close to 0. 

\begin{table}[ph]
\tbl{The exponent of gluon dressing function near the zero momentum 
$\alpha_D$, near the $pa=1$  $\alpha_D'$, the exponent of ghost dressing 
function near the zero momentum $\alpha_G$ and $\alpha_D+2\alpha_G$ in 
$\log U$ type simulation.\vspace*{1pt}}
{\footnotesize
\begin{tabular}{|c|c|rrrr|}
\hline
{} &{} &{} &{} &{}&{}\\[-1.5ex]
$\beta$ & $L$ &{ $\alpha_D$ } & { $\alpha_D'$ } & { $\alpha_G$} &{ $\alpha_D+2\alpha_G$}\\[1ex]
\hline
{6.0 } &{ 32}& { -0.375}  & { 0.302} & { 0.174}&{ -0.03(10)} \\[1ex]
{6.4} & { 48}& { -0.273}  & { 0.288} & { 0.109} &{ -0.06(10)}\\[1ex]
\hline
\end{tabular}\label{tab2} }
\vspace*{-13pt}
\end{table}

ALPHA collaboration\cite{Lue} derived scheme independent running coupling in  
the Schr\"odinger functional method. 
Our data of $\alpha_s$ for $q \geq 1GeV$ is consistent with those of Schr\"odinger functional.
\section{Discussion and Conclusion}
In the lattice simulation of $\widetilde{MOM}$ scheme\cite{orsay1}, the running coupling decreases in the infrared region and 
$\Lambda_{\overline {MS}}=295\pm 20MeV$ was reported. 
Orsay group claims that the non-perturbative gluon condensates
reduces the $\Lambda_{\overline {MS}}$ in the $\widetilde {MOM}$ scheme  from 295MeV to 238MeV and becomes consistent with that in the Schr\"odinger functional method\cite{Orsay2}.  

In the PMS\cite{pms} one assumes that there is an infrared fixed point and one bridges the gap between ultraviolet asymptotic free and the infrared non-perturbative regions.
In our analysis of lattice data we considered 
 optimum coupling constant $y$ at $q=1.97GeV$ that corresponds to $1/a$ of 
$\beta=6.0$ and observed that the PMS in $\widetilde{MOM}$ scheme fits 
the lattice data, $q>0.5GeV$ in the ghost propagator, and $q>1GeV$ in the 
gluon propagator and the running coupling.

In the infrared region, we also analyzed the data via the contour-improved 
perturbation series which is expressed by the Lambert $W$ function\cite{HM}. 
We find that the result is close to 
that of the static quark potential derived by using the perturbative gluon 
condensate dynamics $\tilde \alpha_V$ shown in the analysis of hypothetical 
$\tau$ lepton decay coupling constant\cite{Bro}. Results will be presented elsewhere.

This work is supported by the KEK supercomputing project No.02-82.


\begin{thebibliography}{20}

\bibitem{Gv} V. N. Gribov, {\it Nucl. Phys.} {\bf B139},  1 (1978).

\bibitem{KO} T. Kugo and I. Ojima, {\it Prog. Theor. Phys. Supp.} {\bf 66},  1 (1979).

\bibitem{Ku} T. Kugo, In {\it Int. Symp. on BRS symmetry}, hep-th/9511033.

\bibitem{Zw} D. Zwanziger, {\it Nucl. Phys.} {\bf B364}, 127 (1991),
 {\it Nucl. Phys.} {\bf B412}, 657 (1994).

\bibitem{Zw1} D. Zwanziger, {\it Phys. Lett.}{\bf B257}, 168 (1991).

\bibitem{MN} T. Maskawa, H. Nakajima, {\it Prog. Theor. Phys.} {\bf 60},1526 (1978).

\bibitem{NF} H. Nakajima and S. Furui, {\it Nucl. Phys.} {\bf B} (Proc Suppl.){\bf 63A-C}, 635, 865 (1999) 
hep-lat/9809080,9809081; {\it Nucl. Phys.}{\bf B} (Proc Suppl.){\bf 83-84}, 521(2000) , hep-lat/9909008; In {\it Confinement2000}, p.60, hep-lat/0006002; {\it Nucl. Phys.} {\bf B} (Proc. Suppl.) in press, hep-lat/0208074. 

\bibitem{NFY1} H. Nakajima, S. Furui, A. Yamaguchi, 
{\it Nucl. Phys.} {\bf B} (Proc. Suppl.){\bf 94}, 558 (2001), 
 hep-lat/0010083 v2; {\it Ichep2000 contribution}, hep-lat/0007001.  


\bibitem{FAR} C. S. Fischer, R. Alkofer and H. Reinhardt, hep-ph/0202195.

\bibitem{orsay1} D. Becirevic et al., {\it Phys. Rev.} {\bf D61}, 114508  (2000).

\bibitem{Orsay2} Ph. Boucaud et al., hep-ph/0003020 v2.

\bibitem{Lue} M. L\"uscher, Advanced Lattice QCD, hep-lat/9802029.


\bibitem{pms} P.M. Stevenson, {\it Phys. Rev.} {\bf D23}, 2916 (1981).

\bibitem{vanacol} K. van Acoleyen, {\it Phys. Rev.} {\bf D66}, 125012 (2002).

\bibitem{HM} D.M. Howe and C.J. Maxwell, hep-ph/0204036 v2.

\bibitem{Bro} S.J. Brodsky, S. Menke and C. Merino, hep-ph/0212078 v3.

\end{thebibliography}
\end{document}